\documentclass[pre,twocolumn,showpacs,amsmath,amssymb,floatfix,superscriptaddress]{revtex4}
\usepackage{amsmath}
\usepackage{graphicx}
\usepackage{amssymb}
\usepackage{times}
\begin{document}

\title{Beyond electronics, beyond optics: single circuit parallel computing
with phonons}

\author{Sophia Sklan}
\affiliation{Department of Physics, Massachusetts Institute of Technology, Cambridge, Massachusetts, 02139, United States.}
\author{Jeffrey C. Grossman}
\affiliation{Department of Materials Science and Engineering, Massachusetts Institute of Technology, Cambridge, Massachusetts, 02139, United States. }

\begin{abstract}
Phononic computing \textendash{} the use of (typically thermal) vibrations
for information processing \textendash{} is a nascent technology;
its capabilities are still being discovered. We analyze an alternative
form of phononic computing inspired by optical, rather than electronic,
computing. Using the acoustic Faraday effect, we design a phonon gyrator
and thereby a means of performing computation through the manipulation
of polarization in transverse phonon currents. Moreover, we establish
that our gyrators act as generalized transistors and can construct
digital logic gates. Exploiting the wave nature of phonons and the
similarity of our logic gates, we demonstrate parallel computation
within a single circuit, an effect presently unique to phonons. Finally,
a generic method of designing these parallel circuits is introduced
and used to analyze the feasibility of magneto-acoustic materials
in realizing these circuits.
\end{abstract}


\maketitle
Computers are unquestionably useful, thanks in no small part to their
silicon-based electronic architecture. After all, the size and expense
of vacuum tube computers rendered personal computing impossible. Today,
there are many proposed alternatives to silicon computers. Quantum
computing, for example, promises to perform calculations more efficiently
than any classical computer could \cite{Q Comp}. Moreover, biological
and fuzzy logic computers outperform classical computers in specific
applications but are unlikely to supplant traditional computers \cite{Bio Comp,Fuzzy Comp}.
The creation of phonon (vibrational) equivalents of electronic computer
elements has recently been explored. Phononic (often called thermal)
computing\textquoteright{}s typical application is to use heat from
electronic computers, where waste heat is cheap, plentiful, and wasted
\cite{Phononics,Phn Logic}.

The development of further applications for phonon computers has been
stymied by the difficulty in controlling phonon propagation. This is aggravated
by reliance upon analogies between temperature and voltage, thus taking
inspiration from electronics \cite{Phononics,Phn Logic,Phn Diode,Phn Transistor,Phn Mem}.
Such focus has yielded reliance upon nanostructured materials \cite{CNT Rect,VO Rect}
and 1D, nonlinear chains \cite{Phononics,Phn Logic,Phn Diode,Therm Rect}
to construct circuit elements. While this has resulted in difficulties
with constructing phonon computers, electronic-inspired computation
has also impacted how information is encoded. Prior efforts encoded
information in temperature differences. But when phonons alone are
information carriers, one could take inspiration from optical computing,
encoding information in the polarization of the phonon current. Logical
1 (0) would be vertically (horizontally) polarized transverse phonons.
Logic operations in this computer therefore correspond to modifying
the polarization of phonon currents. Kittel first realized the possibility
of controlling phonon polarization. He discovered the acoustic Faraday
effect (AFE, or acoustic Faraday rotation, it rotates the angle of
polarization) for ultrasonic waves in ferromagnets subject to a uniform
magnetic field \cite{Kittel}. Interest in this application seems
to have ended there. Instead, the focus on magneto-acoustic (MA, also
called magneto-elastic) effects (the influence of a magnetic field
on phonons) has been to discover new mechanisms for MA effects and
how they might aid in characterizing materials \cite{GG Book,AFM Th,TGG Exp}
or controlling magnons \cite{Magnons}.

\begin{figure}
\includegraphics[scale=0.55]{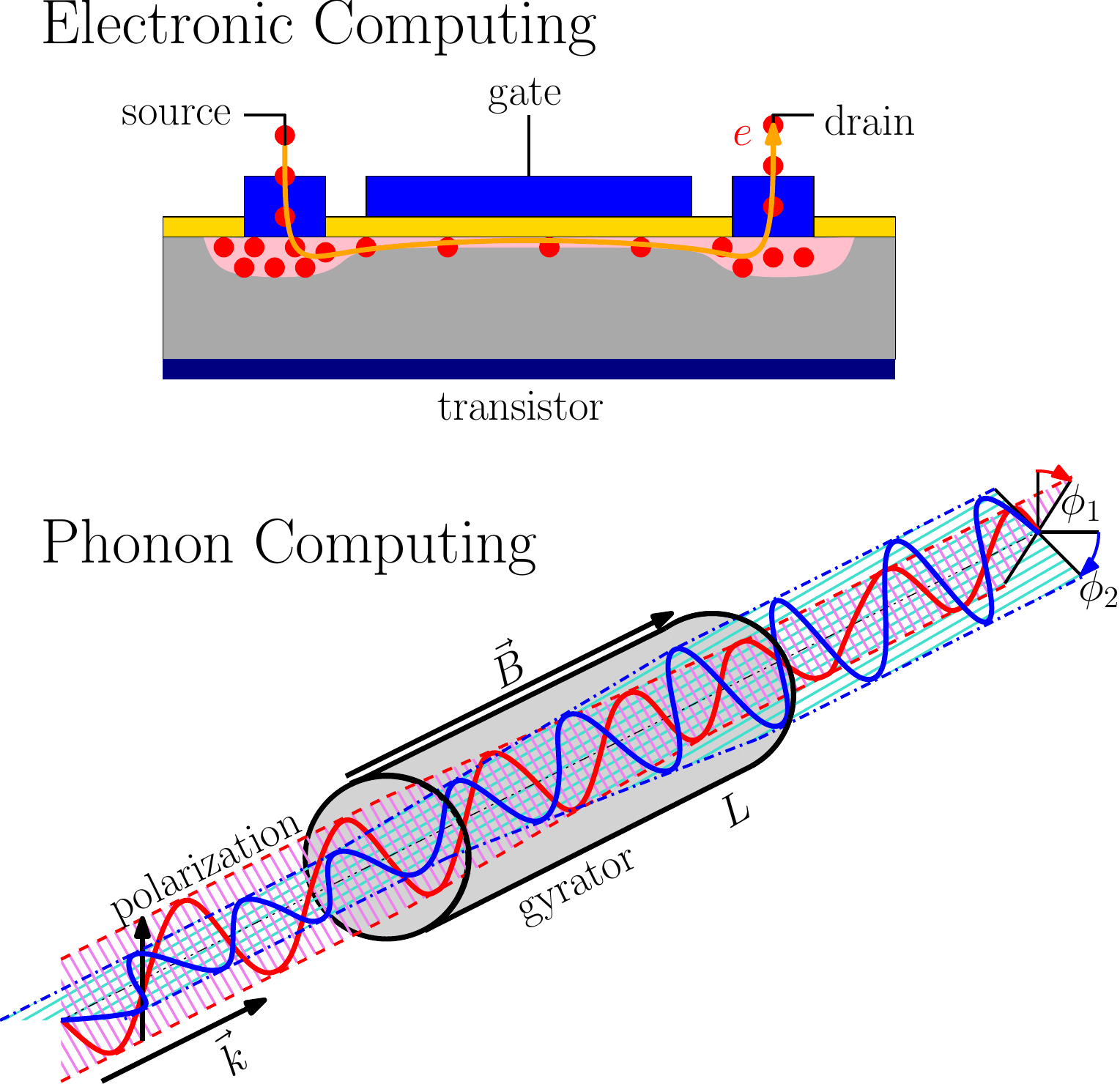}

\caption{(COLOR ONLINE) Comparing electronic and phononic computing. Electronic computing
uses a stream of electrons through a solid-state transistor to control
the magnitude of the voltage at the drain. Conversely, our implementation
of phonon computing has transverse phonons undergo an acoustic Faraday
rotation, controlling the polarization of the transmitted signal.
Note the use two incident phonon currents with different polarizations
and frequencies, each rotated by a different angle while traversing
the gyrator.}
\label{fig:compare}
\end{figure}

Here we consider the construction of logic from MA materials beginning
with the fundamentals - sources, wires, and operators. Attention will
be paid to gyrators, where we find that, in addition to isolators
(the optical analog of diodes), gyrators can create transistors. Logic
gates are constructed from these transistors. We analyze how individual
gates could perform a new form of parallel computing (which we term
rainbow parallelization, RP) (Figure \ref{fig:compare}).

It is helpful to consider where a signal comes from and how it reaches
computing elements. Waste heat would be the ideal source of cheap,
convenient phonons, but it lacks coherence, intensity, and polarization.
At the other extreme, the phonon laser \cite{phonon laser} has good
beam characteristics but is in the early stages of development. For
MA, acoustic transducers are a promising compromise. Piezoelectrics
(e.g. AC-cut quartz) convert electronic signals into polarized phonons
currents, with polarization depending on the crystallographic plane
at the interface \cite{Luthi Book}. At higher frequencies (GHz to
THz, instead of MHz to GHz), femtosecond lasers can be transduced
into picosecond phonon pulses via thin metallic films (e.g. polycrystalline
Al on Zn) \cite{Pico Phn,Pico Phn2}. Once phonons have been produced,
they can be steered via wires. Any material with good lattice thermal
conductivity would do, as this implies that phonons propagate quickly
and retain polarization over long distances (long phonon mean free
path).

While the previous elements are similar to their electrical counterparts,
operators are not. In digital electronics, operators (e.g. transistors)
control voltage levels (logic states) by blocking or admitting electrical
current; whereas operators in polarization-based computing always
admit currents (states are modified by switching polarizations). Given
our basis, a change of state is accomplished by having a gyrator rotate
the polarization by $\pi$/2. Gyrators can be constructed from any
material possessing the AFE (e.g. MA) (Figure \ref{fig:compare}). In the AFE, a longitudinal
magnetic field ($\vec{B}\Vert\vec{k}$, $\vec{k}$ is the phonon wave-vector)
induces circular birefringence (two circularly polarized modes travel
at different speeds). Every linearly polarized mode is composed of
a superposition of circularly polarized modes, which reach the end
of the gyrator at different times. This phase difference between the
circular modes yields a linearly polarized current with a new polarization.
The magnetic field strength determines the magnitude of the gyration
\cite{YIG Exp}. There are many MA materials known \cite{GG Book}, although
yttrium iron garnet (YIG) is a good candidate here as the AFE has
been observed at room temperature \cite{YIG Exp}.

Currently, when gyrators are used, they are paired with a fixed magnet
(often 0.01-10T) to give a predefined gyration. This design has some
uses in computing (e.g. when creating isolators), but gyrators have
other uses. We propose that a phonon current could control the strength
of the magnetic field, allowing the gyrators to operate in a completely
new way. One control mechanism is to transduce a second current, rectifying
and amplifying the resultant signal, and using this signal to allow
or block a fixed current from reaching an electromagnet (Figure S1).
With this sort of controlled magnetic field, gyrators become three
terminal devices that act like generalized transistors. Signals propagate
from one side of the gyrator to the other (source to drain), but the
gating current on the magnet controls the outgoing polarization. Importantly,
treating the gate as the input and the drain as the output, the signal
is amplified (as in electronic transistors). This is an advantage
over isolator-based designs (e.g. some forms of optical computing
\cite{Pht Isolator}), as losses accumulate and hinder detection.

\begin{figure}
\includegraphics[scale=0.85]{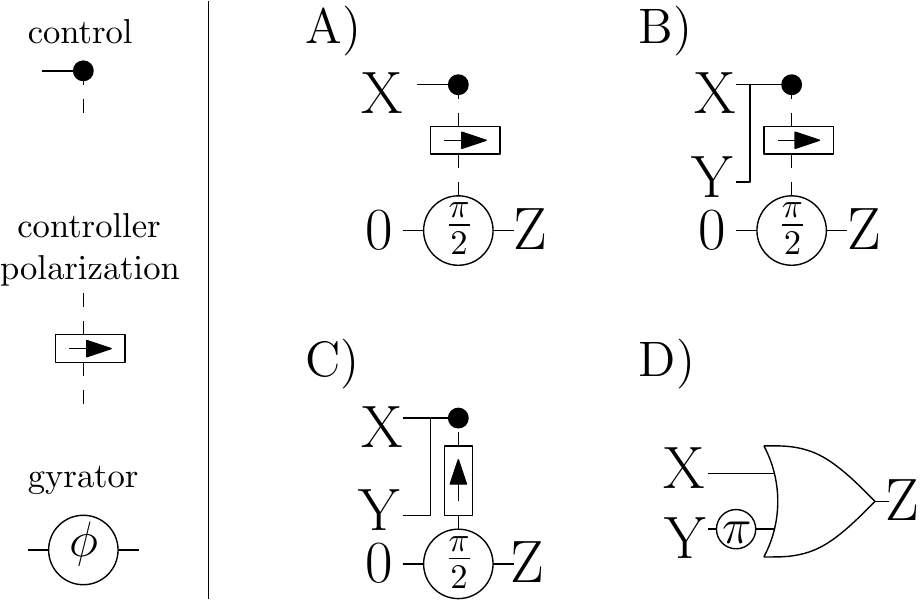}

\caption{Implementing transistor logic with phonon gyrators. Elements are defined
in sidebar. X and Y are inputs, Z is output, and 0 is an input fixed
to logical 0. (a) NOT gate (b) NAND gate (c) OR gate (d) coherent
XOR gate (using OR).}
\label{fig:TTL}
\end{figure}

With these transistors, we can construct basic logic gates. First,
consider a NOT gate. Given our basis, the easiest implementation is
a $\pi$/2 gyrator (0 in gives 1 out, 1 in gives $-$0 out. To remove
the negative sign, which is an overall phase, the output can be fed
into the source and gate of a $\pi$ gyrator, the gyrator activated
by 0). This design is good for changing OR/NOR and AND/NAND, but an
amplifying design is straightforward (Figure \ref{fig:TTL}(a)). Logical 0 is fed
into the source of a gyrator and the signal to be inverted is applied
at the gate. Logical 0 at the gate activates the gyrator, rotating
the source signal by $\pi$/2 , yielding logical 1 at the drain. However,
logical 1 at the gate does not trigger the control, so the gyrator
is not activated and the drain is at logical 0. Now consider two signals
at the gate (Figure \ref{fig:TTL}(b)). When the phonons are in a definite polarization,
two signals will superimpose (no polarization change for identical
inputs and $\pm\pi$/4 polarization for orthogonal signals). Control
activates when this total signal has a horizontally polarized component
greater than a critical value. Thus, if either input is logical 0,
then control will activate and the output will be logical 1. Only
when both inputs are vertical will the output be logical 0. Therefore,
this circuit is NAND. Control could alternatively select logical 1
(Figure \ref{fig:TTL}(c)). Now, the output will only be logical 0 when both inputs
are as well, ergo this circuit is OR. With either NAND/OR and NOT,
we can use De Morgan\textquoteright{}s laws to create any logic gate.
But for coherent phonons, XOR is simple (Figure \ref{fig:TTL}(d)). Essentially,
there is now destructive interference in the OR gate for signals of
identical polarization, so the current can only exceed the critical
amplitude when the signals are orthogonal. And since a $\pm\pi$/4
polarized signal activates control, this OR gate is now an XOR.

\begin{figure}
\includegraphics[scale=0.8]{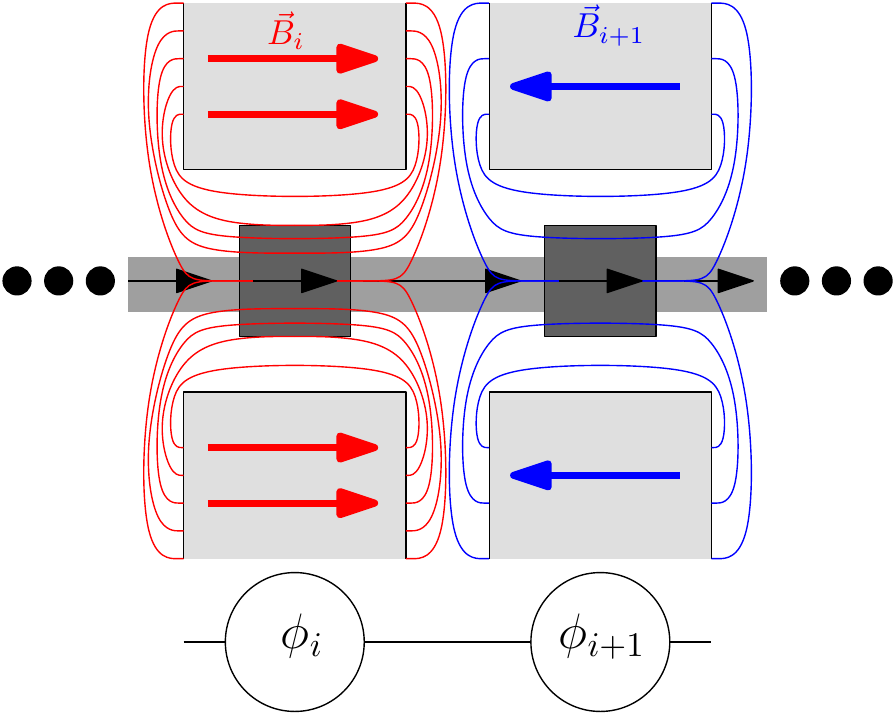}

\caption{(COLOR ONLINE) Multistage gyrator: a concatenation of several gyrators. The lengths
and magnetic field strengths at each stage are tuned to control the
gyrations at multiple frequencies.}
\label{fig:multi}
\end{figure}

While these designs are quite simple, there are other realizations
of logic gates. Exploiting De Morgan\textquoteright{}s laws, we can
construct any operation using only $\pi$/2 gyrations and vertical
control elements (Figure S2). Using gates in fixed arrangement and
selectively (de)activating them, we can perform arbitrary operations
(as in an FPGA). From this, single circuit parallel computing (RP)
follows. Consider multiple signals of different frequencies all entering
the same generic circuit. Independently controlling which gyrators
and control elements work at which frequency, we can independently
select the logical operation for each frequency. Such control is feasible
with multistage gyrators (Figure \ref{fig:multi}) \textendash{} several gyrators
in series where the sum of the gyrations at a given frequency is equal
to the desired gyration \begin{equation}
\Phi(\omega_{(i)})=\underset{(j)}{\sum}L_{(j)}\phi(\omega_{(i)},B_{(j)})\label{eq:1}\end{equation}
$B_{(j)}$ is the magnetic field strength of the $j^{th}$ stage,
$L_{(j)}$ its length, $\omega_{(i)}$ the frequency of the $i^{th}$input,
$\Phi(\omega_{(i)}$) it gyration, and $\phi$ the gyration per unit
length. The number of stages required for a multistage gyrator scales
as $O(N)$ for ungated elements and $O(N^{\alpha})$ ($\alpha\in[1,2]$
and depends upon the design used) for gated elements. These scalings
can be improved by combining stages (allowing fringing fields to come
to bear, slightly complicating the design and discussed in the supplementary
materials). Control of different frequencies can be accomplished by
filtering signals before amplification.

\begin{figure}
\includegraphics[ scale=0.55]{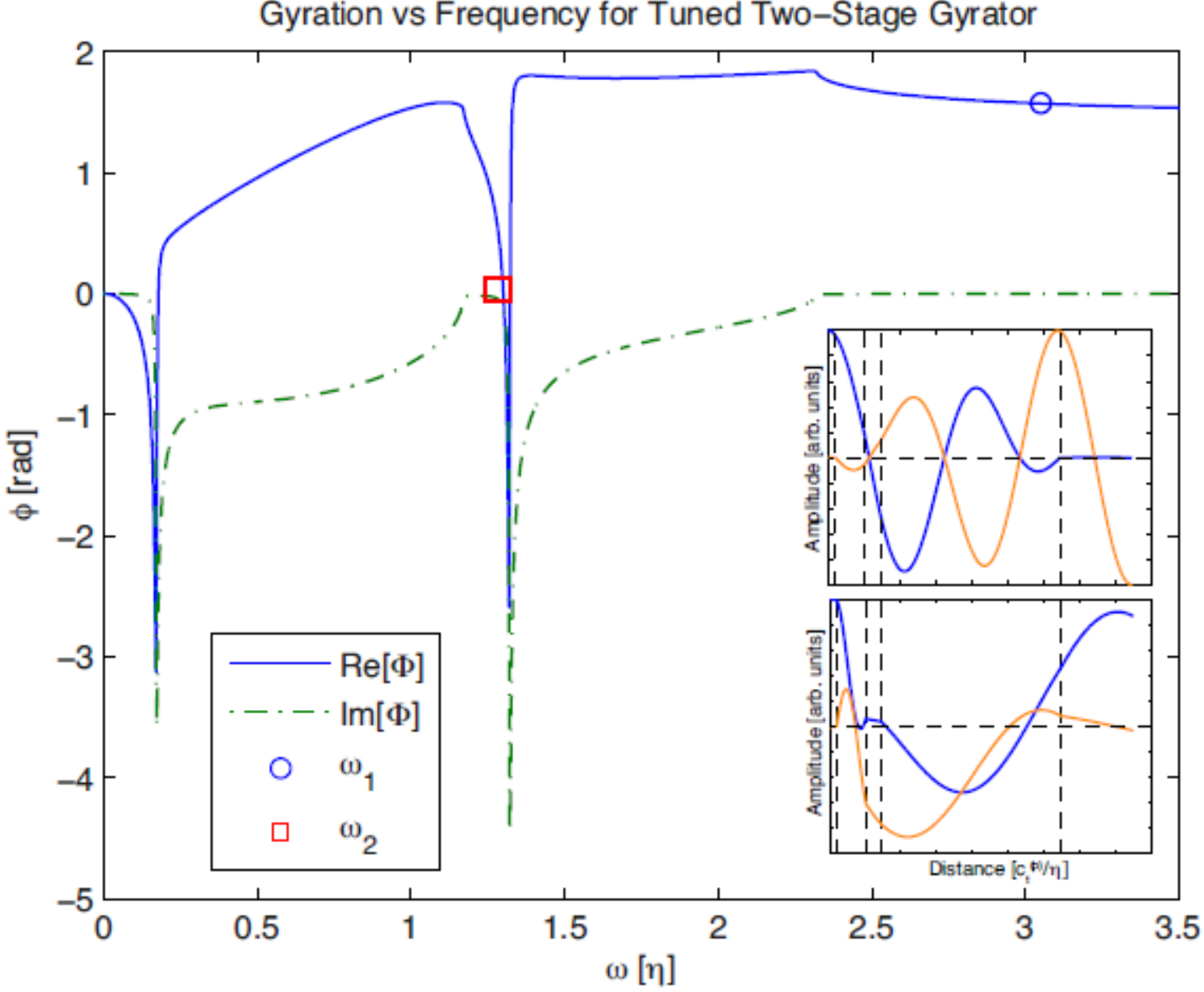}

\caption{(COLOR ONLINE) Rainbow Parallelization: designing a 2-stage gyrator to work at two
frequencies. Total gyration as a function of frequency plotted in
blue. Absorbtion (Im{[}$\Phi${]}) in dashed green. Circled frequencies
have $\Phi=0$ and $\Phi=\pi$/2. $\omega$ is rescaled by $\eta$
(the MA coupling rate). (inset) Real space representation of the waveforms
for the two selected frequencies selected. $L$ is rescaled by $c_{t}^{(p)}/\eta$
($c_{t}^{(p)}$ is the transverse phonon speed). Bottom/top figure
is for lower/higher frequency mode with blue (orange) denoting horizontal
(vertical) polarization. Vertical lines denote boundaries (2nd and
4th intervals are gyrators).}
\label{fig:tuned}
\end{figure}

Assuming uniform magnetic field, the polarization shifts obey Eq.\ref{eq:1},
one can easily find $L_{(i)}(\{B\},\{\Phi\},\{\omega\})$
by inverting $\phi(\omega_{(i)},B_{(j)})$(\{\} denote sets). The
exception to this is when the $\phi(\omega_{(i)},B_{(j)})$ is separable
( $\phi=f(\omega)g(B)$). While this is not generically the case with
phonons, it is for optics (where $\phi=V(\omega)B$) \cite{LFR},
the exceptions principally due to absorption (often in cold gases
\cite{NLFR}). RP is (to our knowledge) presently unique to phonons
(voltage is scalar, so manipulating polarization is impossible in
classical digital circuits), at least at room temperature. To show
the feasibility of this type of design for phonons, we calculate $L_{(i)}(\{B\},\{\Phi\},\{\omega\})$
for a 2-frequency, 2-stage gyrator with no gyration induced at the
lower frequency and a $\pi$/2 gyration induced at the higher frequency.
The gyration $\Phi(\omega;\{B\},\{L\})$ for the multistage gyrator
is plotted in Figure \ref{fig:tuned}. Im{[}$\Phi${]}$<$0 denotes absorbance. Waveforms
at the two frequencies are found analytically in real space (Figure
\ref{fig:tuned} insets). The higher frequency mode has almost no loss ($<10^{-6}$\%),
while the lower has a loss of 19.8\%. Finally, consider the number
of stages possible for a given frequency and magnetic field strength
range. We calculate (see supplementary materials) the maximum number
of channels currently accessible is about 2700. Such a large number
of channels is principally due to the large Q factor of YIG, which
makes it possible to distinguish very small frequency separations.

We have demonstrated a new mode of operation of an acoustic gyrator
and used it to create basic logic gates. By exploiting the wave nature
of phonons and the tunable nonlinearity of MA, we demonstrate that
MA materials can perform RP, an effect that is presently unique to
this computing architecture. This suggests that there might be circumstances
where a phonon computer could outperform an electronic computer. 

This material is based upon work supported by the National Science Foundation Graduate Research Fellowship under Grant No. 1122374. Supporting data and figures relevant to this work are presented online under supplementary material.

\end{document}

\title{Beyond electronics, beyond optics: single circuit parallel computing
with phonons \\ Supplementary Material}

\author{Sophia Sklan}
\affiliation{Department of Physics, Massachusetts Institute of Technology, Cambridge, Massachusetts, 02139, United States.}
\author{Jeffrey C. Grossman}
\affiliation{Department of Materials Science and Engineering, Massachusetts Institute of Technology, Cambridge, Massachusetts, 02139, United States. }

\maketitle

\section{Classes of Magneto-acoustic Materials}

There are several criteria for a material to be able to support the
magnetoacoustic interactions necessary for the acoustic Faraday effect
\cite{GG Book}. First, it must be a crystalline material with at
least $C_{3}$ (3-fold rotation) symmetry. Second, there must be some
intermediate field to couple the phonon modes to the magnetic field.
This is necessary because the Lorentz force acting on a phonon is
generally negligible. The intermediate field can be photons, electronic
charge, or magnons. For photon-phonon coupling, the relevant interactions
arise in low temperature metals, where a magnetic field allows the
propagation of helicon (for uncompensated metals) and doppleron or
Alfven wave (for both compensated and uncompensated metals) modes.
These modes have been studied in pure metals (copper, nickel, etc),
where it is found that dopplerons (which have phase and group velocities
antiparallel to each other) are only supported for very small sets
of wavevectors.

In electron-phonon coupling, the relevant interaction is a coupling
between the octupole moment of the 3d (as in CeAl2), 4f (as in TGG),
or higher order charge distributions and the crystalline electric
field (CEF). These interactions arise because the nuclei in the crystal
are charged, and so distortions of the lattice will induce distortions
of the electronic structure and vice versa. For materials of high
rotational symmetry, the lower order modes of the charge distribution
(monopole, dipole, quadrupole) are less significant. Because these
materials have multiple phonon bands (i.e. acoustic and optical branches),
\cite{TGG Exp,TGG-S} found that they have two types of magnetoacoustic
interactions. In addition to the typical coupling scheme (magnetic
field to electronic charge to crystalline field to phonons) there
is also an indirect coupling, where magnetoelastic effects in one
branch can induce them in other branches due to phonon-phonon couplings.
Additionally, there is a second class of electron-phonon interactions
capable of supporting magnetoacoustic effects. This is doppler-shifted
cyclotron resonance (DSCR), which occurs under similar circumstances
to helicon-phonon and doppleron-phonon couplings (i.e. low temperature
pure metals).

Lastly, in magnon-phonon coupling, the dependence of the exchange
interactions in the spin Hamiltonian upon the location of the nuclei
implies a dependence of the magnon bands upon the vibrational distortion
of the lattice. This is essentially an ionic magnetostriction effect.
It is found in both ferromagnets (for example YIG \cite{Kittel}),
antiferromagnets or flopped antiferromagnets (as in Cr2O3 \cite{AFM Th}),
or paramagnets doped with magnetic impurities (single molecule magnets,
as in KMgF3+Ni \cite{KMgF3+Ni-S}). It is an open question if phonon-phonon
effects can induce a second type of magnetoacousic interaction for
magnon-phonon coupling schemes (the conclusions that we draw here
still work qualitatively if these effects are present, but the quantitative
description would need to be modified). Since the acoustic Faraday
effect is most extensively studied at room temperature in magnon-phonon
coupling schemes (in particular in YIG), we shall YIG for our magnetoacoustic
element.

\section{Control Mechanisms}

There are two basic classes of magnetic control. Control can either
be direct or indirect. In direct control, the magnetic field is modified
by the phonon current itself, and in indirect control the phonon current
transduces some intermediate signal which can control the magnetic
field. Direct control can be performed by a ferromagnet (see Fig.
S1(a)). When a phonon current is incident upon the magnet, a portion
can be absorbed or converted into localized modes that heat up the
magnet. This increase in temperature will result in increased disorder
of the magntization within the crystal, forming domains and weakening
the total magnetization. Since the spontaneous magnetization is temperature
dependent, the resultant magnetic field will be modified after a negligible
delay (since changes in field propagate at the speed of light). If
the temperature achieved is never so great as to demagnetize the entire
magnet, there will be always be a net magnetization in a fixed direction,
which means that upon cooling the down the direction of the total
magnetization will remain fixed (thus keeping our gyrators acting
like gyrators and not polarizers). While this approach is clearly
the simplest to engineer, there are some limitations to direct control.
First, if the thermal conductivity and specific heat of the magnet
are not (approximately) linear within the range of temperatures achieved,
then it is exceedingly difficult to control the strength of the magnetic
field. What's more, the heating of the magnet depends upon the total
thermal current, rather than any particular polarization or frequency.
While polarizers can eliminate transverse polarizations (and even
frequeny intervals, when pairs of orthogonal polarizers are used),
there is no easy way to eliminate the longitudinal modes. While we
can normally safely ignore these modes (as in every other element,
they are completely decoupled), here controlled operation would require
us to either account for or eliminate any longitudinal effects. Lastly,
because the phonons are being converted into magnetic disorder, they
are being destroyed in this process and cannot be used in other circuit
elements.

Indirect control gets around many of these disadvantages. In the control
scheme given here, there are three stages (Fig. S1(b), stages shown
in insets). First, the phonon current is sent through a piezoelectric
material (or some other linear, polarization sensitive transducer).
The piezoelectric is cut such that the faces of the crystal correspond
to the polarization vectors of the basis of transverse modes. The
transverse modes will distort the crystal, inducing a fluctuating
electrization which will produce an oscillating voltage gradient.
This voltage gradient will be directed parallel to the polarization
vector of the phonon current and have magnitude linearly dependent
upon the phonon amplitude, that is $\partial_{i}V\propto\epsilon_{iz}$
(where $\epsilon_{iz}$ is the strain component in the ith direction
of a mode propagating along z). Placing wires at the faces of piezoelectric
will result in a very small AC voltage that depends solely upon a
given phonon mode (i.e. $\Delta V\propto\hat{n}_{i}\epsilon_{iz}$,
where $\hat{n}$ is the normal vector of the crystal surface). Because
the energy extracted is so small, the phonon current leaves the piezoelectric
essentially without loss. At the second stage, this voltage is converted
into a finite, constant voltage (removing all non-transient time dependence
is necessary to keep the magnetic field from fluctuating). This conversion
is accomplished by using a concatenation of voltage multiplying rectifiers.
The number of rectifiers used will increase the output voltage at
this stage, so arbitrary voltages can be achieved in principle. This
voltage could be used to drive the current in an electromagnet (stage
three), which would give a magnetic field that is approximately linear
in the amplitude of a polarized phonon current (linearity is maintained
so long as the total magnetic field remains well below the maximum
possible). Strict linearity, however, can be a drawback in building
digital circuits. For example, in the construction of an OR gate it
is preferable to have signals of strength $\epsilon$ and 2$\epsilon$
give identical fields (corresponding to inputs of X=1, Y=0 and X=1
Y=1, respectively). To achive this sort of digital control, the voltage
produced at the end of stage two is used to control the gate of an
electronic transistor (see Fig. S1(c)). Since the current produced is
the one that flows from source to drain, the transistor output is
(approximately) a digital response to the gate voltage. The amplitude
dependence of these three types of control regimes is summarized in
Fig. S1(d). Since large magnetic fields are typically required to achieve
strong magnetoacoustic effects (typically in the range of .5 to 10
T \cite{AFM Th,YIG Exp}, sometimes fields of up to 20 T have been
used \cite{TGG Exp}), the electromagnet will have to be quite powerful.
Bitter electromagnets, for example, have been constructed to produce
continuous fields of at least 35 T (superconducting magnets can exceed
this, but not at room temperature) \cite{Bitter}, far higher than
any the fields required by our system.

\includegraphics[scale=0.4]{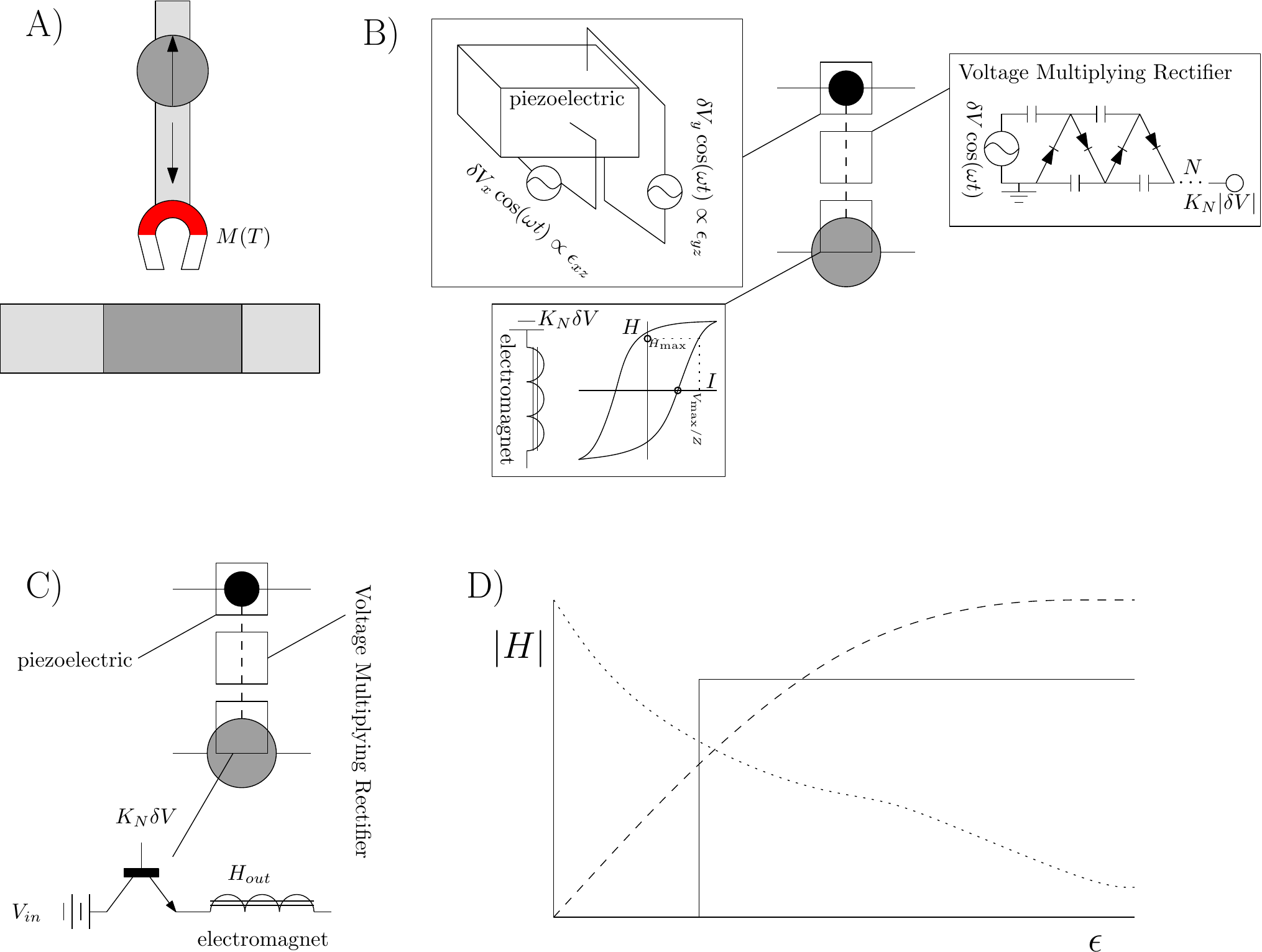}

FIG S1: Magnetic control designs. (a) Direct control by phonons. (b)
Linear indirect control. (c) Digital indirect control. (d) Magnetic
field as a function of amplitude. The dotted line denotes direct control.
The dashed line denotes linear indirect control. The solid line denotes
digital indirect control.

\section{Similarity of Logic Gates}

To determine the minimum number of gates required to construct all
the two-terminal logic functions, we specialize to only using $\pi/2$
gyrations and veritcal control. By explicit construction (see Fig. S2), we see that
we require three control elements (one for each input, one for their
superposition), and seven gyrators (three controlled, four fixed).
Given that control elements are always paired to gyrators, this gives
a total of seven elements to perform all sixteen logic operations.

\includegraphics[scale=0.5]{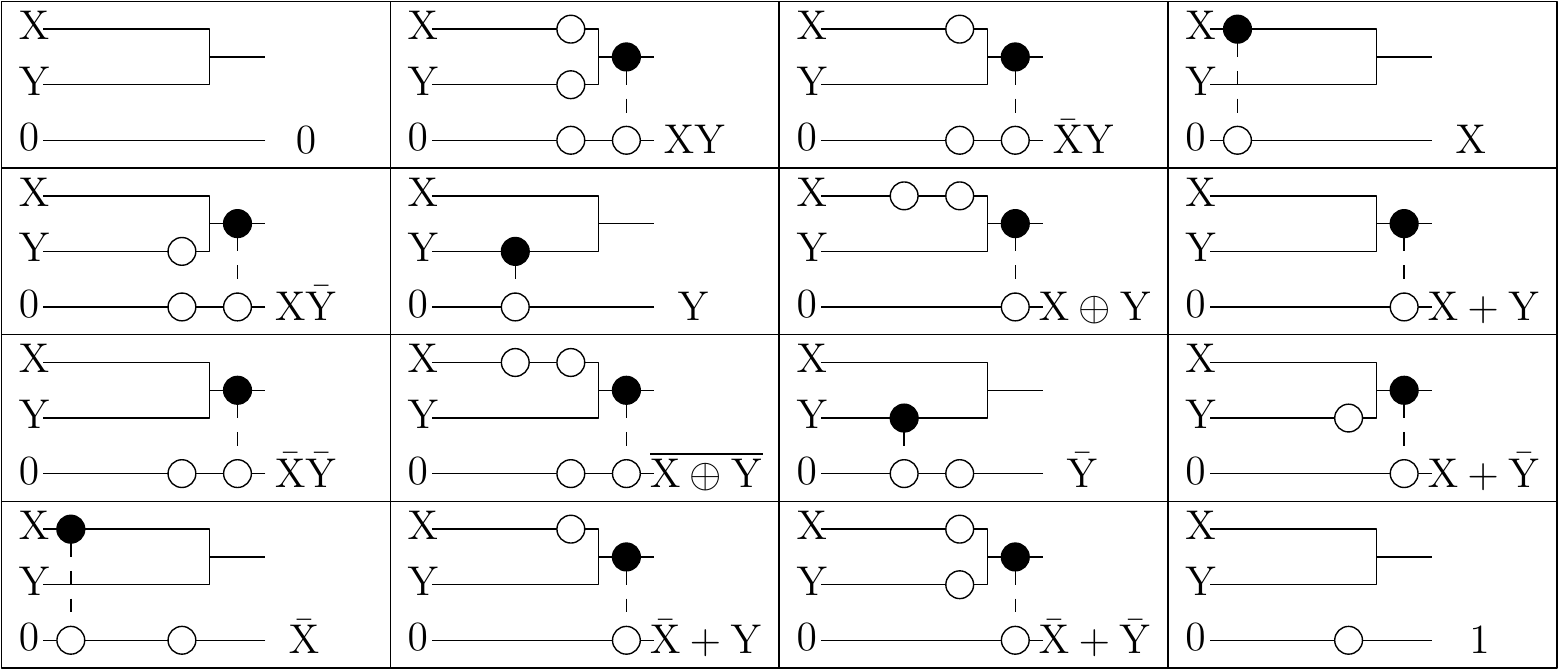}

FIG S2: Alternative transistor logic using only vertical control
and 90 degree gyrations. Implementations of all 16 two-terminal logic
functions.\\

\section{Incorporation of Inhomogeneity into Multi-Stage Gyrator Design}

The Faraday rotation induced by a uniform magnetic field is straightforward
to calculate. When the magnetoacoustic interaction is due to magnon-phonon
resonance, it takes the form \begin{eqnarray*}
\Phi & = & \int dz\frac{k_{+}-k_{-}}{2}\\
 & \approx & V\int dz\frac{\omega^{2}}{\omega^{2}-\Omega^{2}}\\
 & = & L\phi\end{eqnarray*}
 where $\Phi$ is the total rotation, $V$ is a constant (the analog
of the Verdet constant), $\omega$ the phonon frequency, $\Omega$
the effective magnon frequency, $z$ the runs over the thickness of
the gyrator along the direction of propagation ($L)$, and $\phi$
is the gyration per unit length \cite{GG Book}. In a strong, uniform
magnetic field ($\vec{B}\equiv B\hat{z}$), $\hbar\Omega\approx\gamma|B|$,
where $\gamma$ is the gyromagnetic ratio ($g_{s}\mu_{B}$ for spins).
When the magnetic field is not uniform (which can occur for multistage
gyrators), calculating the Faraday modes is more complicated. We assume
a gradual inhomogeneity (that is, only including the effects of varying
the magnon modes themselves, ignoring the deviation of the coupled
mode Hamiltonian from the Faraday regime). Consider the inhomogeneous
magnetic field $B=-\frac{1}{2}B^{\prime}(z)(\vec{x}+\vec{y})+B(z)\hat{z},$
where $B^{\prime}(z)=\frac{dB(z)}{dz}\ll B(z)$ and the term proportional
to the derivative of the field is necessary to satisfy $\nabla\cdot B=0.$
This gives a vector potential $A=-\frac{1}{2}B(z)(y\hat{x}-x\hat{y}),$
thereby maintaining the orbital Zeeman term proportional to $B(z)L_{z}$.
The spin Zeeman term, which is more important in this context, is
proportional to $B\cdot S$. The electronic degrees of freedom will
arrange themselves properly so that $B(z)S_{z}$ can be evaluated
unambiguously at fixed $z$, i.e. that $\omega_{nq}^{(m)}\to\omega_{nq}^{(m)(0)}(z),$
($\omega_{nq}^{(m)(0)}$ is the magnon frequency for the nth branch/polarization
and wavevector $q$, calculated to 0th order in transverse component)
which can be integrated over in determining the Faraday rotation.
The terms perpendicular to $\hat{z}$, on the other hand, will perturb
the magnon Hamiltonian, shifting the frequencies. We treat these additional
terms as a pertubation upon $\omega_{nq}^{(m)}(z),$ keeping terms
less than $O(\alpha^{3})$, as we did in our previous derivation.
The perturbation Hamiltonian, in Holstein-Primakoff coordinates is:
\[
\Sigma_{q}B^{\prime}(z)\sqrt{2S}[x^{-}(-q)\alpha(q)+x^{+}(-q)\alpha^{\dagger}(q)]\]
 where \begin{eqnarray*}
x^{\pm}(q) & = & -i[\frac{\partial}{\partial q_{x}}\pm i\frac{\partial}{\partial q_{y}}]\\
 & = & -i\sqrt{\frac{3}{8\pi}}\mathcal{F}[rY_{1}^{\mp1}(\hat{r})]\end{eqnarray*}
 ($\mathcal{F}$ is the Fourier transform operator, $r$ is distance,
and $Y_{l}^{m}$ are the spherical harmonics), which (to lowest order)
can be considered a constant so far as spin degrees of freedom are
concerned. The first order correction vanishes trivially, and the
second order is \[
\frac{2S[B^{\prime}(z)]^{2}}{\hbar^{2}\omega_{nq}^{(m)(0)}(z)}\left[|\langle x_{-q}^{-}\rangle|^{2}-|\langle x_{-q}^{+}\rangle|^{2}\right],\]
which is generically nonzero. 

\section{Design of a Multistage Gyrator}

While the use of inhomogeneous magnetic fields makes for a more elegant
form of single circuit parallelization (in principle, only a single
magnetoacoustic material is required, with the magnetic field programmable
by a set of control elements), it does make design more difficult.
For uniform fields, the relation \begin{eqnarray*}
\Phi(\omega_{i}) & =\underset{(j)}{\sum} & L_{(j)}\phi(\omega_{(i)},B_{(j)})\end{eqnarray*}
 is clearly satisfied, meaning that we can invert the $\phi$ matrix
to determine the necessary length. To demonstrate how this design
works, we consider the problem of designing a two frequency circuit
which flips one mode and passes the other. To calculate the exact
Faraday rotation we use $k_{\pm}^{2}=k_{0}^{2}(1+\frac{\eta}{\omega_{s}\pm\omega-\eta}$)
where $c_{t}^{(p)}k_{0}=\omega$ ($c_{t}^{(p)}$ is the transverse
speed of sound), $\omega_{s}$ the magnon frequency, and $\hbar\eta=4|G_{44}|^{2}S^{3}/M$
is magnetoelastic coupling energy ($G_{44}$ is the coupling constant,
$S$ is spin, and $M$ is the effective mass). The Faraday rotation
equation given previously is based upon the assumption that $k_{+}-k_{-}=(k_{+}^{2}-k_{-}^{2})/2k_{0}$
(i.e. $k_{+}+k_{-}\approx2k_{0}$). While this approximation is generally
useful, it does not give an easy way to calculate the absorption of
the incident wave.

We work in units such that the equations simplify, i.e. $\Omega\equiv\omega_{s}-\eta\approx\gamma B-\eta+i\Gamma\equiv B+i\Gamma$,
and rescale all the variables ($\omega\to\omega/\eta$ , $B\to B/\eta$,
$\Gamma\to\Gamma/\eta$, $k\to c_{t}^{(p)}k/\eta$, $L\to\eta L/c_{t}^{(p)}$,
$\phi\to\phi$) giving the relation $k_{\pm}^{2}=\omega^{2}(1+\frac{1}{B+i\Gamma\pm\omega})$
For clarity, we keep the magnon lifetime ($\tau_{s}=1/\Gamma$) long,
setting $\Gamma$ =0.0023. We select $B_{1}$=1.32 and $B_{2}$=0.17,
thus making it unlikely that our values happen to simplify the problem
(we also want the magnetic field values reasonably well separated,
as significant losses appear in the range Re$[\omega]=[B,B+1]$).
For $\Phi_{2}=\pi/2$, it is best to select a frequency far from resonance,
so we set $\omega_{1}$=3.05. To make $\Phi_{1}=0$, it is necessary
for it to be close to resonance with a magnon mode (this is implied
by the shape of the Faraday rotation curve). Thus, $\omega_{2}^{2}=1.3^{2}-\Gamma^{2}$.
Solving for the lengths then gives $L_{1}=$ 0.403 and $L_{2}$=2.51.
A plot of the total gyration as a function of frequency quickly confirms
that this device acts as predicted. To translate this plot into the
real space wavefunction, we assume impedance matching at the boundaries
(i.e. ignoring losses due to reflection). This allows us to isolate
the losses due to the finite lifetime, which we find to be 3.81$\times10^{-7}$\%
at $\omega_{1}$ and 19.8\% at $\omega_{2}$.

\section{The Maximum Number of Channels Currently Accessible}

The number of possible channels for a multi-stage gyrator is difficult
to determine. In principle, an infinite number of stages could support
an infinite number of channels, but since the complexity of the device
increases as the number of stages increases. Moreover, there are limitations
to the frequencies and field strengths that are feasible. As such,
only a small subset of the channels physically accessible. Since the
strength of the magnetic field can be controlled experimentally so
long as the electromagnets used are below their saturation magnetization,
we are limited more by the resolution of phonon modes. We estimate
the spectral resolution of the phonon modes using the Q-factor. To
make sure that the frequencies are distinguisable, we assume $\omega_{N}+\frac{1}{2}\Delta\omega_{N}=\omega_{N+1}-\frac{1}{2}\Delta\omega_{N+1}$
where $\omega_{N}$ is the Nth frequency used and $\Delta\omega$
is the bandwidth. Essentially, this relation implies that frequencies
are separated by their average bandwidth. Plugging in the definition
of $Q_{N}=\omega_{N}/\Delta\omega_{N}$, we find that \[
\omega_{N}/\omega_{0}=\Pi_{n=0}^{N}\frac{2Q_{n}+1}{2Q_{n}-1}\le(\frac{2Q_{\min}+1}{2Q_{\min}-1})^{N}.\]
 Therefore \[
N\ge\ln(\frac{\omega_{N}}{\omega_{0}})/\ln(\frac{2Q_{\min}+1}{2Q_{\min}-1})\approx Q_{\min}\ln\frac{\omega_{N}}{\omega_{0}}\]
 (the error in this approximation is $<$1\% for Q$>$10). Setting $\omega_{N}=\omega_{\max},\omega_{0}=\omega_{\min}$
gives the number of distinguishable frequencies accessible. Experimentally,
the majority of MA experiments concentrate in the range $\omega/2\pi\in$
10MHz to GHz \cite{GG Book}. For magnetic fields of $\le$10T, however,
the cyclotron frequency (i.e. the magnon frequency) is 152 MHz. We
therefore lack the ability to easily tune gyrations above this frequency.
Q for YIG is typically quite large in this frequency, values between
$10^{4}$ and $10^{7}$ are regularly found for frequencies between
10MHz and 1GHz ($Q\propto1/\omega$) \cite{Q-factor}. However, when
the intensity of the phonon current is too great (1mW is the rule
of thumb in the 500MHz-1GHz range), the magnon coupling is no longer
linear and the Q factor degrades (the phonons lose energy to magnons,
which quickly decay). This degradation is typically of order unity,
but we shall take a conservative estimate of Q as $10^{3}$. This
gives $N=$2700. This could be improved even further by countering
the reduction of the Q factor, as well as extending the range of accessible
frequencies. Conversely, given that this many channels would make
fabrication dramatically more complicated, the a portion of the channels
could be sacrificed (e.g. reducing the maximum strength magnetic field
to a more readily achievable value).